DARK MATTER AND DARK ENERGY AS EFFECTS OF QUANTUM GRAVITY

Max I. Fomitchev[1]

ABSTRACT

I present a theory of quantum gravity based on the principle of gravitational energy fluctuations. Gravitational energy fluctuations – *gravitons* – are responsible for elastic scattering of subatomic particles. Such scattering corresponds to complimentary force – *graviton scattering force* – arising in gravitational interaction in addition to Newtonian gravity. The strength of the graviton scattering force is proportional to the graviton scattering probability. Unlike Newtonian gravity the graviton scattering force follows the $r^{-1}$ law and dominates the former on cosmological scale in the limit of low orbital accelerations.

Similarly to Modified Newtonian Dynamics (MOND) the quantum gravity accounts for variations in observed *M/L* ratios of diverse stellar systems ranging from dwarf spheroid galaxies to X-ray galaxy clusters without requiring an invisible matter (which is still required by MOND in X-Ray cluster cores). Unlike MOND the presented theory neither violates cornerstone Newton Laws nor suffers from the ambiguity of acceleration frames while enjoying vast experimental evidence usually cited in favor of MOND.

To ascertain the validity of the presented theory I have examined the predictions of quantum gravity for dwarf spheroid, ordinary and giant elliptic galaxies, and X-ray clusters. In all cases quantum gravity yields *M/L* ratios and scaling relations consistent with observations. Moreover, quantum gravity accounts for the tilt of the Fundamental Plane of elliptical galaxies erasing the differences in *M/L* vs. luminosity relations for faint and bright ellipticals, which cannot be easily explained by the standard CDM model.

Lastly, by analyzing the behavior of the gravitational energy fluctuations in the limit of high matter density expected in the early Universe I show that primordial inflation and dark energy (i.e. non-zero cosmological constant) arise as natural effects of quantum gravity in the expanding Universe.

*Subject headings*: cosmology: dark matter – galaxies: kinematics and dynamics – X-rays: galaxies: clusters

---

[1] [1] Email: fomitchev@psu.edu; Postal address: 861 Willard St, State College, PA 16803, USA

## 1. INTRODUCTIOIN

At present time our understanding of gravity is far from being complete: on large scale gravitational dynamics is described by immensely successful theory of general relativity (or its Newtonian limit) while in micro-level particle motion is governed by quantum mechanics; but no bridge yet exists between the two extremes. In the same time cosmology and astrophysics are growing increasingly dependent on research in particle physics and quantum dynamics, especially in conjunction with "dark matter" and "dark energy" theories that are called to explain unexpected observational data. Thus we cannot be sure that our interpretation of motion of celestial bodies by means of classical gravitational theory, which cannot yet be extrapolated down to microscopic level where quantum effects are expected to dominate, is fully accurate.

Not surprisingly, there is a great deal of intellectual effort devoted to research in quantum gravity (see for e.g. Burgess 2003). The view shared by most quantum gravity scientists that at micro-level space-time continuum breaks down into a discreet and even disjointed "foam" is very hard to reconcile with the principles of general relativity simply because the latter requires a space-time continuum! Therefore I adopt a different view, which preserves space-time continuum at micro-level yet relies on principles of quantum dynamics to describe particle motion on microscopic level.

## 2. PRINCIPLES OF QUANTUM GRAVITY

Classical gravity prescribes that a massive body produces smooth and featureless $r^{-1}$ potential well, the notion, which is a very naïve in the realm of quantum physics. More realistic quantum mechanical view calls for a macroscopically smooth potential field to produce energy fluctuations at microscopic level. From the stochastic point of view, however, fluctuations are expected to arise when a complex system of bodies is involved (which is true of any real-world situation). Then what is the *intrinsic* shape of the potential well produced by a *fundamentally simple* body in a space devoid of other bodies and their corresponding fields? To arrive at a realistic picture I postulate that the intrinsic shape of the potential well produced by a fundamentally simple body is a spherically symmetric $r^{-1}$ potential *combined* with the energy of intrinsic fluctuations expressed by a spherically symmetric wave of the amplitude diminishing as $r^{-1}$ and wavelength increasing as $r$:

$$\varphi(r) \equiv \frac{GM}{r} + \frac{\widetilde{U}_0}{r}\sin\left(\frac{\widetilde{l}_0 r}{l_1 r + l_2}\right) \quad (1)$$

Because the second constituent of the potential $\varphi(r)$ is very small the resulting field of a macroscopic body is well approximated by smooth $r^{-1}$ potential employed in classical dynamics. However, effects of the wave constituent will be important when motion of microscopic bodies is concerned.

Having postulated in equation (1) the intrinsic potential of a fundamentally simple body, which radial profile is depicted on fig. 1, we can calculate the field of a complex body as a superposition of a large number of intrinsic potentials – fig. 2.

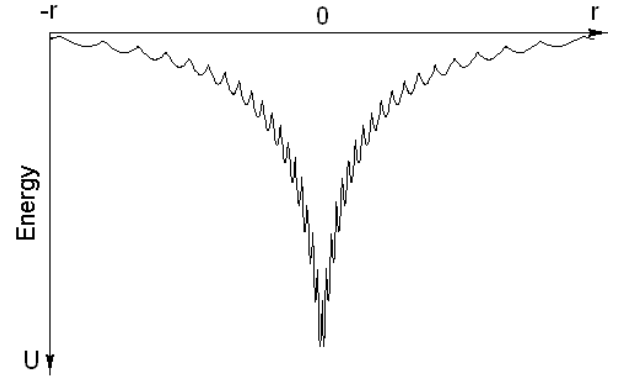

Fig. 1. Radial profile of the intrinsic quantum gravitational potential.

By examining the result of the superposition of the large number of intrinsic potentials one can easily notice circular groove structure that becomes more and more regular (i.e. round) with distance from the center of mass. In the same time individual stochastic fluctuations or *gravitons*, depicted on the overlay image on fig. 2 are clearly seen on microscopic level. The resulting stochastic fluctuations correspond to elongated ellipsoids arranged in groves and oriented tangentially to groves. On the grand scale the fluctuations pervade the entire Universe and form a mesh-like structure of peaks and pits corresponding to superposition (or intersection) of groves produced by neighboring bodies.

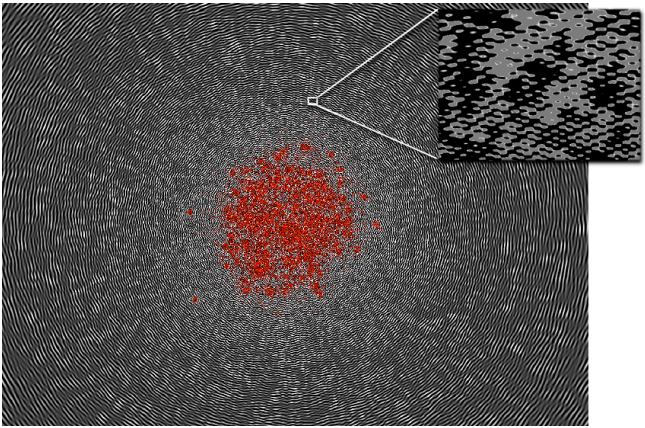

Fig. 2. Potential of a complex body modeled as a superposition of intrinsic potentials. Small detail is artificially enhanced while the global $r^{-1}$ bias is removed. Note circular groove structure that becomes more and more round with distance from the center of mass. Overlay corresponds to a magnified area of the potential depicting individual small-scale stochastic fluctuations – *gravitons* – arising in the field.

When a motion of microscopic particles is concerned such particles no longer travel in empty space but instead move through the universal field of gravitational energy fluctuations and interact with gravitons, which are organized in groves closer to massive bodies and form quasi-static stochastic mesh far from gravitating masses. Gravitational energy fluctuations will form stochastic potential barriers on particle trajectory thus causing reflection from or tunneling through the potential barrier. Because gravitational energy fluctuations are localized (even when aligned in groves) and do not form a continuous barrier the particles will also scatter from gravitons in addition to reflection and tunneling. Therefore within the framework of quantum gravity motion of a particle corresponds to a process of scattering from / tunneling through the field of gravitons globally biased by the classical Newtonian (or relativistic) gravitational potential.

Naturally, for a free particle the vector of the momentary displacement will coincide with the direction of the greatest tunneling probability. Therefore to account for direction the velocity vector must correspond to a "skew" in the particle's intrinsic kinetic energy profile – fig. 3.

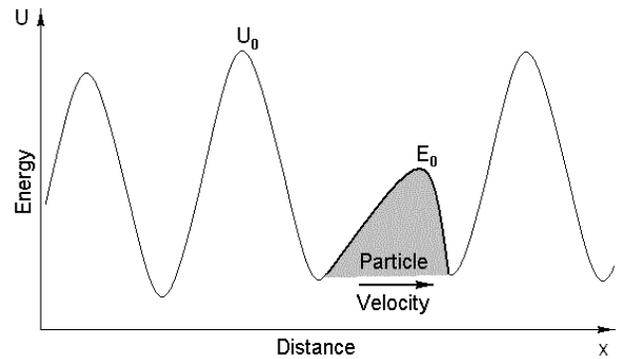

Fig. 3. Particle with energy $E$ confined between gravitational field fluctuations with energy $U$.

The greater particle's intrinsic energy is – the greater is the probability of tunneling especially in the preferred direction determined by the skew in the particle's intrinsic energy field, thus the greater is the particle's velocity. Interesting, but the concept of tunneling within the framework of quantum gravity implies maximum velocity corresponding to tunneling probability $D = 1$. Supposing that maximum velocity corresponds to the speed of light $c$ the particle *tunneling velocity* $V$ can be defined as

$$V = cD \qquad (2)$$

In the case of a complex body corresponding to a bound system of particles the macroscopic velocity of the body will be expressed by a different formula than the equation (2). First of all, the stability of a macroscopic body requires binding energy $E_B$ to be greater than the energies of compounding particles ($E_B \gg E$) or the system will break apart in no time – fig. 4. For example the total energy of nucleons is in the order of 1 GeV and is much greater than the energy of "free" quarks, which is thought to be on the order of 5-10 MeV for u- and d-quarks. Thus, the constituents of a complex body are subjected to interaction with gravitational field fluctuations (i.e. gravitons) within the limits set by the total binding energy of the body.

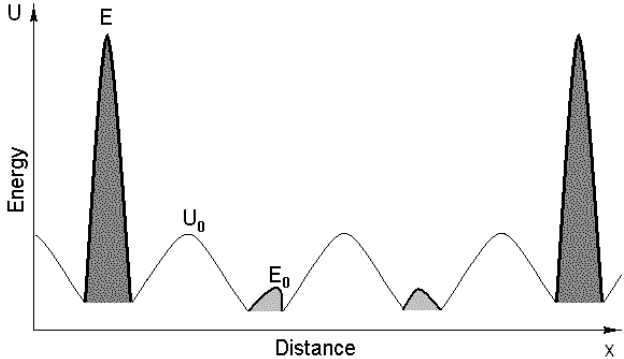

Fig. 4. Energy diagram of a complex body. Individual particles with energy $E$ are confined within a potential well with energy $E_B$ binding the particles together.

## 3. GRAVITON SCATTERING FORCE

The most interesting question to ask is what effects on orbital motion of celestial bodies the proposed theory of quantum gravity can possibly have? To answer this question let us consider a circular orbital motion of a complex body in a Newtonian limit, which is classically expressed in terms of the Newtonian gravitational acceleration $g_N$

$$g_N = \frac{GM}{r^2} \qquad (3)$$

When considering a scenario involving the field of gravitational energy fluctuations in the form as depicted on fig. 2 it is clear that an orbiting body will experience an additional force due to scattering of particles from gravitons. The circular grove arrangement of gravitons as well as their elongated shape favors scattering *towards* the center of mass on circular orbits and reproaching trajectories and scattering *away* from the center of mass on approaching trajectories – fig. 5.

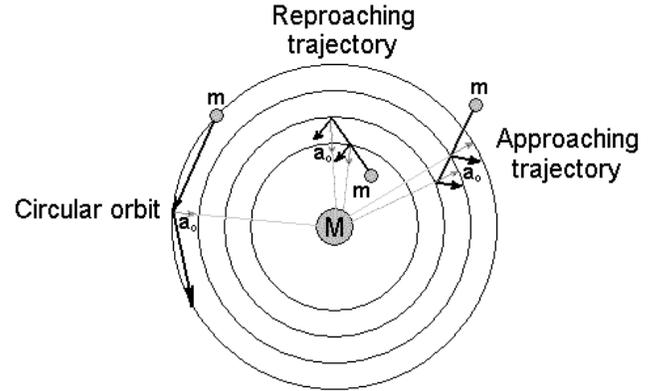

Fig. 5. Direction of graviton scattering force on circular orbit, approaching and reproaching trajectories.

Thus in the case of circular or near-circular orbital motion an additional central acceleration $a_Q$ will be acting on the orbiting body:

$$a = g_N + a_Q \qquad (4)$$

Additional acceleration $a_0$ can be calculated from the change of impulse $dp/dt$ as follows:

$$a_0 = \frac{1}{m}\frac{dp}{dt} = v\frac{v}{l}\sigma(1-D) \qquad (5)$$

where $\sigma$ is a scattering probability, $D$ – tunneling probability through gravitons, and $1/l$ – is linear density of gravitons (which is equivalent to the characteristic graviton size except for a constant scaling factor), and $m$ is the total mass of the body.

To understand the validity of equation (5) one must consider the process of motion on microscopic level: a complex macroscopic body corresponding to a bound system of microscopic particles is moving through the field of gravitons arranged in groves; the microscopic particles scatter from gravitons in the preferred direction either towards the center of mass or away from it depending on the trajectory of the body; some particles, however, tunnel through the fluctuations. Thus a change in the impulse vector of the macroscopic body occurs. Assuming that the energy $E$ of microscopic particles comprising the body is much less than the energy of fluctuations $U$ (i.e. $E << U$) the scattering is going to be elastic with no change in the impulse magnitude but only in direction. The $v/l$ term in equation (5) accounts for the rate of scattering, which is directly proportional to the body velocity and inversely proportional to the linear density of gravitons.

Non-relativistic scattering probability $\sigma$ of particles is inversely proportional to the square of the particle's impulse (Blokhintsev 1983):

$$\sigma \propto \frac{4\pi\hbar^2}{p^2} \quad (6)$$

Assuming that the internal energy density of a graviton can be approximated by a rectangular potential barrier, tunneling probability $D$ for the case of $U > E$ can be calculated as (Yariv 1982):

$$D = \left[1 + \frac{U^2}{4E(U-E)}\sinh^2\left(\frac{l}{\hbar}\sqrt{2m_0(U-E)}\right)\right]^{-1} \quad (7)$$

which in the case of $\frac{l}{\hbar}\sqrt{2m_0(U-E)} << 1$ can be approximated as

$$D = \frac{1}{1 + (U^2 l^2/\hbar^2)(2E/m_0)} = \frac{\gamma^{-2} 2E/m_0}{1 + \gamma^{-2} 2E/m_0} \quad (8)$$

and

$$1 - D = \frac{1}{1 + \gamma^{-2} 2E/m_0} \quad (9)$$

where $\gamma$ has a dimension of velocity:

$$\gamma^2 = \frac{U^2 l^2}{\hbar^2} \quad (10)$$

Now, remembering that $U >> E$ it is safe to assume that the intrinsic motion of microscopic particles within the macroscopic body is negligible and the comprising particles are moved "helplessly" by collisions with gravitons within the volume of space defined by the binding energy of the macroscopic system. In such conditions the comprising particles are forced to tunnel through gravitons only when they reach an edge of the potential barrier defining the macroscopic system. At that moment the microscopic particle assumes velocity of the macroscopic system due to the elastic collision with a larger potential barrier. After being endowed with velocity of the macroscopic system the comprising particles scatter on gravitons. Therefore, for non-relativistic velocities $E = \frac{m_0 v^2}{2}$ and we can rewrite the equations (8) and (9) as

$$D = \frac{\gamma^{-2} v^2}{1 + \gamma^{-2} v^2} \approx 1 \quad (11)$$

$$1 - D = \frac{1}{1 + \gamma^{-2} v^2} \approx \frac{\gamma^2}{v^2} \quad (12)$$

The approximation implies that $\gamma^{-2} v^2 << 1$, which is necessary to ensure that the gravitation scattering is small.

Thus the final equation for the acceleration produced by the gravitational energy fluctuations on circular orbit is

$$a_Q = \frac{4\pi\hbar^2 \gamma^2}{m_0^2 v^2 l} \quad (13)$$

Note that the circular groove alignment of gravitons and the change in direction of the graviton scattering force depending on the approaching or reproaching trajectories implies that all non-circular orbits will tend to relax to circular ones due to the corrective effect of the graviton scattering force. This quantum gravitational preference for orbits with low eccentricity can be responsible for deep isothermality of stellar systems that are pressure supported (i. e. elliptic and spheroid galaxies, intercluster gas).

On the scope of galactic orbits a hint of this behavior is observed in Virgo cluster. Rubin, Waterman & Kenney (1999) report that elliptic galaxies in Virgo (i.e. old cluster population) exhibit a tight Gaussian distribution of observed velocities near 1100 km s$^{-1}$, while spirals with normal rotation curves (i.e. younger population) have velocities evenly distributed in the range from –400 km s$^{-1}$ to 2600 km s$^{-1}$. Curiously, spirals with disturbed rotation curves display velocity distribution almost exactly matching the Gaussian distribution for ellipticals. Because of the match in velocity distribution I conclude that the graviton scattering force is acting on young spiral galaxies moving on radial orbits effectively reducing orbital eccentricity and thus causing disruption of rotation. It is doubtful that tidal interaction is responsible for this effect as it is unlikely to produce correlation in velocity dispersion with older elliptical galaxies, which in the framework of quantum gravity should already be on relaxed circular orbits.

## 4. PROPERTIES OF FUNCTIONS OF $U$ AND $l$

Equation (13) states that the graviton scattering force depends the unknown functions $U$ and $l$, which describe the energy and characteristic size of gravitons. With the quantum gravitational potential of a point source defined by equation (1) the energy $U$ and the characteristic size $l$ of the gravitational energy fluctuations produced by a system of point sources will depend on the number (and possibly density) of the sources or the *total mass* of the system.

To determine the functions $U$ and $l$ a numerical simulation was conducted involving $10^5$-$10^6$ quantum gravitational sources arranged in a sphere with density distributions ranging from $\rho(r) = const$ to $\rho(r) \propto r^{-1}$ and $\rho(r) \propto r^{-2}$. The resulting functions $U$ and $l$ were sampled at various points in space to uncover patterns. The simulation revealed that for spherically symmetric distribution the functions of $U$ and $l$ are independent of density and obey the following laws:

$$l(r) = l_0 r \tag{14}$$

$$U(r) = \frac{U_0}{1 + r/r_0} \approx \frac{U_0}{r}, r \gg r_0 \tag{15}$$

$$U_0 \approx k\sqrt{M} \tag{16}$$

The functions (14) and (15) (which imply that $U(r)\, l(r) \approx U_0\, l_0$ regardless of $r$) are a direct consequence of the postulate (1) for the potential of a quantum gravitational point source. Additionally the equation (15) prescribes that in the core of the system the function $U$ varies more slowly with $r^{-1}$ thus avoiding the infinity at $r = 0$ otherwise implied by the $r^{-1}$ law. Also, $r_0$ is a diminishing function of density:

$$r_0 = f(\rho) \tag{17}$$

What is especially interesting about the function of $U_0$ is that the mass in equation (16) is the *total mass* of the system and *not* the mass enclosed within the radius $r$. This extremely important property of the $U_0$ sets graviton scattering force apart from Newtonian gravity, which for spherically symmetrical distribution depends only on the mass enclosed within the radius under consideration.

Armed with the knowledge of the behavior of functions $U$ and $l$ the equation for the acceleration induced by the graviton scattering force on circular orbit can be rewritten as:

$$a_Q = \frac{4\pi \hbar^2 k^2 M_{total} l_0}{m_0^2 v^2 r} \tag{18}$$

The equation (18) states that the contribution to the orbital acceleration by the graviton scattering force obeys the $r^{-1}$ law in contrast to the Newtonian gravity component, which follows the conventional $r^{-2}$ law.

Based on the equation (18) the total circular acceleration can be expressed as:

$$a = \frac{GM(r)}{r^2}\left(1 + \frac{\lambda}{a}\right) \quad (19)$$

where $\lambda$ has a dimension of acceleration:

$$\lambda = \lambda_0 \frac{M_{total}}{M(r)} \quad (20)$$

and

$$\lambda_0 \equiv \frac{4\pi k^2 l_0}{G m_0^2} \quad (21)$$

In the case of a point mass (or highly localized mass distribution) or for large radii ($r \rightarrow \infty$) $\lambda$ in equation (19) is constant: $\lambda = \lambda_0$.

## 5. THEORETICAL M/L RATIO

From the equation for circular orbital acceleration (19) it is possible to derive a theoretical M/L ratio for a quantum gravitational system. Supposing that the M/L ratio is equivalent to the ratio of the dynamical mass – $M_{dyn} = ar^2/G$ – calculated from the from the virial theorem presuming that the system is in a state of equilibrium to the observed luminous mass, then

$$\frac{M}{L} \propto \frac{M_{dyn}}{M_{obs}} = \frac{ar^2}{GM} \quad (22)$$

And from the equation (19) it follows that:

$$\frac{M}{L} \propto \Upsilon_0 \left(1 + \frac{\lambda}{a}\right) \quad (23)$$

where $\Upsilon_0$ is a 'standard' M/L ratio ($\Upsilon_0 \approx 5$ in B-band or $\Upsilon_0 \approx 1$ in V-band).

Also, in the limit $\lambda \gg a$

$$a = \frac{1}{r}\sqrt{GM(r)\lambda} \quad (24)$$

and

$$\frac{M}{L} = \Upsilon_0 r \sqrt{\frac{\lambda}{GM(r)}} = \frac{\Upsilon_0 r}{M(r)}\sqrt{\frac{\lambda_0 M_{total}}{G}} \quad (25)$$

Assuming that the distribution of matter in the system under consideration is arranged around of a compact core of radius $r_c$ containing the bulk of the system mass ($M(r>r_c) \approx M_{total}$) and the light traces mass (i.e. $L \propto M$) the equation for the theoretical M/L ratio becomes

$$\frac{M}{L} \propto \frac{r}{\sqrt{L}} \quad (26)$$

The equation (26) is also true for dispersed systems at large radii.

In the central part of a dispersed system where $M(r<r_c) \ll M_{total}$, however, a different relation describes the theoretical mass-to-light ratio:

$$\frac{M}{L} \propto \frac{r}{L} \quad (27)$$

Within the very core of a dispersed system where the energy of gravitons $U$ does not decay with radius the theoretical M/L ratio is

$$\frac{M}{L} \propto \frac{r^2}{L} \quad (28)$$

The three scenarios for mass-to-light ratios described by equations (26)-(28) follow from the first principles of quantum gravity. In the following sections I am going to apply the derived M/L ratios to various stellar systems to verify the predictions of quantum gravity.

## 7. FLAT ROTATIONAL CURVES OF SPIRAL GALAXIES AND TULLY-FISHER RELATION

For stellar systems with substantial central mass concentrations (e. g. spiral galaxies including S0-type) and for elliptic galaxies that are not exceedingly sparse and faint (i.e. all ellipticals but dSph galaxies) quantum gravity predicts Tully-Fisher law describing the correlation between maximum circular velocity $v_{max}$ and total luminosity for rotation-supported systems (i.e. spiral galaxies) and Faber-Jackson relationship between central velocity dispersions $\sigma_v$ and total luminosity for pressure-supported systems such as elliptical galaxies.

Tully-Fisher relation can be derived as follows. One can see from the equation (19) that the graviton scattering force will dominate Newtonian gravity in the case of non-zero $\lambda$ and low orbital accelerations. Thus for $a \ll \lambda$ the circular orbital acceleration is

$$a(r) = \frac{v^2(r)}{r} \approx \frac{1}{r}\sqrt{GM(r)\lambda} \qquad (29)$$

The equation (29) implies an asymptotically flat rotational curve with maximum rotational velocity

$$v_{max} = (G\lambda_0 M_{total})^{0.25} \qquad (30)$$

The equation (30) is essentially a Tully-Fisher law $v_{max} \propto M^{0.25}$ for rotational curves of spiral galaxies, which follows from the first principles of quantum gravity.

Also, from the equation (19) follows that in the core of the system the rotational curve is dominated by Newtonian gravity because there $a \gg \lambda$.

## 8. COMPARISON OF QUANTUM GRAVITY TO MOND

There is another theory – Modified Newtonian Dynamics (MOND) – that predicts flat rotational curves of spiral galaxies and Tully-Fisher / Faber-Jackson laws for maximum rotational velocity and line-of-site velocity dispersions vs. total optical luminosity relations. Modified Newtonian Dynamics was pioneered by Milgrom over two decades ago in attempt to eliminate the dark matter problem (Milgrom 1983a,b,c; see also Milgrom 1998). While MOND is a phenomenologically-driven theory, which incorporates the observed Tully-Fisher relation as *one of it postulates*, MOND enjoys tremendous success in explaining the observed effects of 'dark matter' in diverse stellar systems ranging from globular clusters to galaxy clusters (e.g. Sanders & McGaugh 2002). Surprisingly, MOND works even for those stellar systems, for which the theory was not originally designed (see e.g. $M_{ICM}$ vs. $T_x$ relationship for X-ray clusters discussed in Aguirre, Schaye & Quataert 2001). Despite all this success MOND is rejected by a majority of physicists as an alternative to dark matter. The reasons for reluctance of the scientific community to embrace MOND are the lack of basis for MOND built with *conventional* physics; profound theoretical difficulties such as violation of the Equivalence Principle and the 2nd Newton's Law (Bekenstein & Milgrom 1984); the lack of relativistic extension (see Milgrom 2001 and references there in); and the ambiguity of acceleration frames, which makes almost impossible to derive concrete predictions with MOND on cosmological scale (Scott et al. 2001). Yet because the quality of current observational data cannot definitely rule-out MOND, the debate over this theory continues.

The reason I decided to mention MOND in this paper is that there are certain similarities between MOND and the presented theory of quantum gravity. For instance in the limit of low accelerations ($g_N \ll a_0$) MOND states that

$$a = \sqrt{g_N a_0} \qquad (31)$$

which is strikingly similar (to within an factor of +1) to the prediction of quantum gravity:

$$a = g_N (1 + \frac{\lambda}{a}) \qquad (32)$$

In the spirit of this parallel

$$\lambda \approx a_0 \qquad (33)$$

The main differences between the two theories are as follows:

1) The quantum gravity predicts an extra force *addition* to Newtonian gravity, while MOND postulates a *modification* of Newtonian gravity;

2) The quantum gravity is based on conventional physics such as quantum mechanics while MOND is not;

3) Flat rotational curves and the Tully-Fisher law follow from the basic principles of quantum gravity while MOND postulates these relations;

4) The graviton scattering force in quantum gravity depends on the *total* mass of the system, while the modification of gravity in MOND depends on the mass enclosed within the radius of consideration;

5) Although $\lambda \approx a_0$, $\lambda$ is not constant in quantum gravity, while $a_0$ is constant in MOND.

The last two points make big difference when predictions of both theories are applied to clusters of galaxies. Most notably MOND breaks down in X-ray cluster cores where it fails to reproduce the observed X-ray emitting gas temperature profiles (Aguirre et al. 2001) and cannot account for the observed *M/L* ratio without requiring large amounts of unseen matter (Gerbal et al. 1992). Obviously aware of the problem for MOND in X-ray cluster cores, Milgrom (1998) stipulates that there must baryonic matter in X-ray cluster cores (e.g. 'warm' hydrogen gas), which is not yet detected. Although present-date cooling flow deposition rates are insufficient to account for the missing matter in the cluster cores, Milgrom suggests that the rates could have been higher in the past. Ironically, a EUV-excess in Virgo (Lieu et al. 1996a), Coma (Lieu et al. 1996b) cluster was recently reported adding oil to the fire of the X-ray cluster core controversy. Unfortunately, no firm conclusion can be drawn yet about the reported UV excess as the opinions are split among researchers with regard to the origin of the UV emission: while Bonamente & Lieu (2000) firmly believe that the UV excess if produced by large amounts of warm gas much needed for the rescue of MOND, the other group (Bowyer, Korpela & Berghöfer 2001) advocates a strongly opposing point and even challenge the excess EUV findings of the former group (e.g. Berghöfer & Bowyer 2002).

Regardless of the EUV excess findings quantum gravity *does not* require large amounts of unseen matter in X-ray cluster cores. Although should the discussed deposits of 'warm' hydrogen gas really exist in X-ray cluster cores their presence or absence is not critical for the predictions of quantum gravity as the unlike MOND the quantum gravity theory is concerned with the total mass rather than the mass in the core.

Lastly I would like to remind that Abramovici & Vager 1986 have shown that the Second Newton Law $F=ma$ holds down to accelerations as low as $10^{-9}$ cm s$^{-2}$, which is clearly below the critical MOND acceleration $a_0 \sim 10^{-8}$ cm s$^{-2}$ and therefore at odds with MOND.

Because of the similarity of the two theories, however, which amounts to the effective action at low accelerations, quantum gravity benefits from the vast amount of evidence usually cited in favor of MOND (see e.g. Milgrom 2003, Scarpa 2003, Sanders & McGaugh 2002, Sanders 2001) while avoiding the noted shortcoming of MOND.

## 9. DWARF SPHEROID GALAXIES: DARK MATTER CONTENT AND VELOCITY DISPERSIONS

Measurements of central velocity dispersions in dwarf spheroid galaxies – dSphs – reveal dynamical masses of dSph systems far greater than their observed luminous masses and resulting in *M/L* ratios between 10 and 100 (Mateo 1994). Although there is a considerable uncertainty in the estimates of *M/L* ratios for dSph galaxies (in part due to their extreme faintness and large angular sizes), it is clear that the stellar content alone cannot account for the observed velocity dispersions.

Besides large dark matter content dwarf spheroid galaxies are characterized by a central velocity dispersions being in the range of 5-12

km s$^{-1}$ while their luminosities span about two orders of magnitude (Gallagher & Wyse 1994). Thus unlike spiral or elliptical galaxies dSph systems show no *strong* correlation between the velocity dispersions and optical luminosities when taken individually. Although velocity dispersions of dSph galaxies are plotted versus luminosity along with other dwarf systems such as dIr and dE galaxies they fall on the same $\sigma_v \propto L^{0.25}$ trend very tightly (Dekel & Woo 2003). Also, there exists great uncertainty in measuring total optical luminosities of dwarf spheroid galaxies because of their large angular size and extreme faintness, which along with other effects such as tidal interaction and hydrogen cooling at $T \gtrsim 10^4$ may contribute to the leveling off of the $\sigma_v$ vs. $L$ relation for dwarf systems in general (Dekel & Woo 2003).

In the realm of quantum gravity dwarf spheroid systems are expected to show signs of large amounts of dark matter. From equation (26) it follows that the less luminosity there is in a dwarf the higher its *M/L* ratio will be. This trend is in fact very well established (e.g. Mateo et al. 1993). Also Mateo et al. (1993) finds that *M/L* ratios for dwarf spheroid galaxies in B-band have a constant bias of $\Upsilon_B \sim 5$, which could be indicative of the contribution of the Newtonian gravity predicted by the equation (23).

What else quantum gravity is telling us about dwarf systems is that these objects will remain *bound* by the graviton scattering force for a wide range of radii even when these systems are subjected to tidal disruption. Perhaps the tidal disruption is responsible for breaks in dSph stellar density distributions, which can be fitted by two-component King profiles (e.g. Pryor & Kormendy 1990). Tidal interaction will tend to send stars in outer regions of dwarf systems on higher orbits bound by the graviton scattering force rather than simply tear stars away. The net effect of such interaction is that the system stellar density is diluted the surface brightness is decreased with time. Support for this argument can be drawn from the observed correlation between dwarf metallicity and surface brightness, which reveals that metal poor systems are much fainter than their metal-rich counterparts (Caldwell et al. 1992). Presuming that lower metallicity is indicative of greater age, the observed relationship implies that the stellar densities of older dSph systems are in fact diluted and the system geometry is expanded and smeared by tidal interaction proportionally to the age of the system.

## 10. ELLIPTIC GALAXIES

### 10.1. Dark Matter Content

Dark matter content of elliptical galaxies of all sizes has become a subject of intense debate for the most part due to the surprising findings of Romanowsky et al. (2003) indicating an unexpected 'dearth' of dark matter in 'ordinary' elliptic galaxies with characteristic luminosities in the order of $2.2 \times 10^{10}$ $L_{B,\odot}$ (in B-band solar units for Hubble constant $H_0 = 70$ km s$^{-1}$ Mpc$^{-1}$). While both giant and dwarf ellipticals display signs of vast amounts of 'dark matter' in their outer *M/L* ratios – whether in the case of strong gravitational lensing (e.g. Keeton 2001) or X-ray emission profiles (e.g. Loewenstein & White 1999) for giant systems or in abnormally high stellar velocity dispersions in nearby dwarfs (e.g. Mateo 1997) – according to Romanowsky et al. (2003) intermediate-size elliptic galaxies contain very little or no dark matter at all. This is a disturbing finding for CDM theorists.

This apparent discrepancy is easily understood within the framework of quantum gravity. The equation for orbital acceleration (19) predicts the Keplerian decline in circular velocity profiles with radius to level off at the maximum circular velocity $v_{max}$ dictated by the graviton scattering force. This leveling off of the circular velocity profile is attributed to the presence of dark matter when the galaxy is analyzed within the context of Newtonian gravity. In the realm of quantum gravity magnitude of the maximum circular velocity depends on the total system mass as prescribed by the equation (30).

Thus for massive ellipticals with luminosities ~ $10^{13}$ $L_\odot$ and maximum circular velocities $v_{max}$ ~200-300 km s$^{-1}$ the contribution of the graviton scattering force to the orbital acceleration is significant and is due to the large stellar mass of these galaxies – see equation (18). Therefore at large radii graviton scattering force exceeds Newtonian gravity many times, which results in high observed $M/L$ ratios.

Going down three orders of magnitude in mass / luminosity we arrive at 'ordinary' ellipticals with characteristic luminosities ~$10^{10}$ $L_\odot$. According to (21) the maximum circular velocity $v_{max}$ for ordinary elliptical galaxies will be in the range 40-60 km s$^{-1}$, which is comparable to the Newtonian circular velocities appropriate for their masses, hence the low dark matter content.

Going down another three orders of magnitude in mass / luminosity we arrive at dwarf systems with luminosities ~$10^7$ $L_\odot$. The equation (30) puts maximum circular velocity at ~10 km s$^{-1}$. This time contribution of the graviton scattering force is many times the magnitude of the force of Newtonian gravity, and the observed 'dark matter content' is high.

Similar reasoning for $M/L$ ratios in elliptical galaxies of various luminosities can be given in MONDian spirit: equation (29) relates orbital acceleration to the $M/L$ ratio as $M/L \propto \lambda/a$. For systems with low orbital accelerations (e.g. dwarf systems and outer rims of giant galaxies) the $M/L$ ratios are high, while for galaxies with high orbital accelerations (e.g. 'ordinary' ellipticals) the resulting $M/L$ is not much greater than unity.

### 10.2. Fundamental plane

Another interesting consequence of quantum gravity is that the theoretical $M/L$ ratio (31) provides a natural explanation for the tilt of the Fundamental Plane of elliptical galaxies for faint dwarf and bright giant ellipticals alike. These two types of systems follow two vastly different trends that cannot be reconciled with the help of the virial theorem alone (see e.g. Busarello et al. 1997) without resorting to rather ad hoc systematic change in $M/L$ with luminosity or departure from homology.

It is generally accepted that faint systems such as dwarf spheroid and dwarf elliptical galaxies trace the following trend (Dekel & Silk 1986; Peterson & Caldwell 1993)

$$M/L \propto L^{-0.4} \qquad (34)$$

The equation (34) is consistent with the prediction of quantum gravity, which follows from the equation (26) for the theoretical $M/L$ ratio:

$$M/L \propto L^{-0.5} \qquad (35)$$

On the other hand bright elliptical galaxies follow a totally different relation (Gerhard & Kronawitter 2000):

$$M/L \propto L^{0.37} \qquad (36)$$

Gerhard & Kronawitter (2000) also obtain the following scaling relationship for system size vs. luminosity:

$$r \propto L^{0.81} \qquad (37)$$

Combining the reported relation (37) with the theoretical $M/L$ ratio (26) yields

$$M/L \propto L^{0.31} \qquad (38)$$

In the same time combining the relation (36) the theoretical $M/L$ ratio (26) produces

$$r \propto L^{0.87} \qquad (39)$$

Thus the observed scaling relations for giant ellipticals fit nicely the expectations of quantum gravity.

Even more support for the quantum gravitational explanation of the tilt of the fundamental plane of elliptic galaxies can be drawn from the detailed study conducted by Bernardi et al. (2003) of rather substantial sample of 9000 early type galaxies obtained from the Sloan Digital Sky Survey. For their sample Bernardi et al. (2003) reports:

$$M/L \propto L^{0.14\pm0.02} \qquad (40)$$

$$M/L \propto R^{0.33\pm0.09} \qquad (41)$$

The latter implies that

$$R \propto L^{1.0\pm0.6} \qquad (42)$$

Once again the reported results are consistent with the *M/L* ratio predicted by quantum in equation (26) within the stated uncertainties with the favorable *R* vs. *L* law for the sample given by the following relation:

$$R \propto L^{1.3} \qquad (43)$$

Thus the theoretical *M/L* ratio (26) predicted by quantum gravity successfully describes the tilt of the Fundamental Plane for dwarf, ordinary and giant elliptical galaxies erasing the discrepancy that exists when analyzing the phenomenon using the accepted CDM model, which requires ad hoc effects such as evolution or systematic variation of *M/L* with luminosity to explain the difference in tilt.

## 11. X-RAY CLUSTERS

*11.1 X-Ray Temperature And M/L Profiles*

When considered in the framework of quantum gravity X-Ray galaxy clusters fall roughly in the same category as dSph systems. This similarity is not accidental and is caused by the fact that both types of systems correspond to relatively sparse and geometrically extended stellar systems where accelerations are measured *inside* the mass distribution rather than outside (which is the case for outer edges of spiral and elliptical galaxies). The discussion of X-ray clusters is especially important because MOND cannot account for high *M/L* ratios detected in X-ray cluster cores without requiring vast amounts of unseen matter (Gerbal et al. 1992) and further fails to reproduce flat X-ray emitting gas temperature profiles observed in galaxy clusters (Aguirre et al. 2001). In the same time MOND successfully explains *global M/L* ratios for galaxy clusters and even predicts accurate mass-temperature relation for X-ray clusters (Aguirre et al. 2001):

$$M_J \approx 4.6 \times 10^{12} M_{solar} \left(\frac{kT}{\text{keV}}\right)^2 \qquad (44)$$

In other words predictions of MOND imply that the bulk of the cluster mass is concentrated in the cluster core, e.g. in the form of warm gas as suggested by Milgrom. The latter expectation has been a reason for a major debate regarding the existence of large quantities of warm hydrogen gas in galaxy cluster cores, which was already discusses in §8 of this paper. Although a firm conclusion has not yet been reached in the cluster EUV excess debate, the outcome of the discussion has little bearing on quantum gravity. As I have pointed out in §4 of this paper the exact mass distribution does not matter very much: it is the *total mass* that counts. Thus in the context of X-ray clusters quantum gravity behaves similar to MOND except that the *total mass* of the cluster figures in the equation for the additional acceleration produced by the graviton scattering force thus avoiding the difficulty that MOND experiences in cluster cores. In other words in the realm of quantum gravity *M/L* ratio for X-ray clusters will remain more or less flat with increasing radius. The latter expectation is consistent with the flat *M/L* profiles reported for a sample of 59 nearby clusters analyzed by Katgert, Biviano & Mazure (2003).

The expectation of a flat *M/L* profile for a cluster, which is a result of the dependence of the contribution of the graviton scattering force on total mass of the system, comes in extremely handy in explaining the flatness of the observed X-ray emitting gas temperature profiles with radius. Aguirre et al. (2001) point out that for gas in hydrostatic equilibrium MOND predicts that

$$T(r) \propto \sqrt{GM(r)a_0} \propto \sqrt{M(r)} \qquad (45)$$

In the realm of quantum gravity the equation (45) becomes

$$T(r) \propto \sqrt{GM(r)\lambda} \propto \sqrt{M_{total}} \qquad (46)$$

Thus in quantum gravity $T(r)$ is independent of radius and thus consistent with the reported isothermality of the X-ray emitting gas, which is practically independent of radius.

*11.2 Scaling Relations*

Another advantage of quantum gravity is that it reproduces the observed scaling relations for clusters that differ from the expectations of self-similarity in self-gravitating systems described by Newtonian dynamics.

Namely, self-similarity implies that (see e.g. Ettori et al. 2003)

$$L \propto T^2 \qquad (47)$$

$$M_{dyn} \propto T^{1.5} \qquad (48)$$

In the same time Ettori et al. (2003) reports

$$L \propto M_{dyn}^{1.88 \pm 0.42} \qquad (49)$$

$$L \propto T^{3.7 \pm 1.0} \qquad (50)$$

$$M_{dyn} \propto T^{1.98 \pm 0.30} \qquad (51)$$

I specifically used the designation $M_{dyn}$ for total dynamical mass inferred from the virial theorem as $M_{dyn} = v^2 r/G$, to avoid confusion with the total mass $M_{total}$ employed in the equations for quantum gravity. There is, however, a simple link between the two quantities: since the virial theorem implies that $M_{dyn} \propto v^2$ and in quantum gravity $v \propto \sqrt[4]{M_{total}}$ then

$$M_{dyn} \propto \sqrt{M_{total}} \qquad (52)$$

Substituting $M_{dyn}$ with $M_{total}$ in the observed scaling relations (49)-(51) one obtains that

$$T \propto M_{total}^{0.5} \qquad (53)$$

$$L \propto M_{total} \qquad (54)$$

The obtained relations (53) and (54) are fully consistent with the predictions of quantum gravity.

As far as $\sigma_v$-$T$ relation is concerned there seems to be a consensus in the literature that $\sigma_v \propto T^{0.6}$ (see e.g. Xue & Wu 2000). The reported $\sigma_v$-$T$ relation is consistent with the expectation of quantum gravity that $\sigma_v \propto T^{0.5}$, which follows from the facts that $\sigma_v \propto M^{0.25}$ and $T \propto M_{total}^{0.5}$.

## 12. DARK ENERGY AS EFFECT OF EXPANSION

As individual gravitating bodies produce gravitational field fluctuations Universe as a whole becomes a source of energy fluctuations (i.e. gravitons). In expanding universe fluctuations caused by individual bodies will recede thus producing an accelerating drag associated with expansion. By analogy with the equation (5) the expansion-induced acceleration $a_{exp}$ can be expressed as

$$a_{exp} = \frac{v^2}{l}(1-D)\sigma \propto \frac{(1-D)}{l} \qquad (55)$$

where $l$ is a characteristic size of the universal gravitons and $v$ is a particle velocity relative to the flow of expansion (if the rate of expansion at present time is characterized by the Hubble law then the particle velocity $v$ in equation (55) is the particle's peculiar velocity relative to the Hubble flow).

Thus the energy of expansion is

$$U_{exp} \equiv a_{exp} R \propto \frac{(1-D)}{l} R \qquad (56)$$

Because the energy of expansion (56) is non-zero and its effect is repulsive when the Universe expands such energy will act as cosmological term:

$$\Lambda \equiv a_{exp} = \Lambda_0 \frac{1-D}{l} \qquad (57)$$

where $\Lambda_0$ is constant.

## 12. COSMOLOGICAL INFLATION AS EFFECT OF EXPANSION

Since within the framework of quantum gravity the cosmological constant is non-zero due to the effect of the gravitational energy fluctuations, the equation for the rate of the expansion of the Universe can be written as

$$\frac{\ddot{a}}{a} = -\frac{4}{3}\pi G(\rho + p) + \frac{\Lambda}{3} \quad (58)$$

To analyze the expansion of the early Universe it is necessary to determine the dependence of the characteristic size $l$ of the gravitons on the size of the Universe $R$, i.e. the function $l(R)$. Clearly, when the density of the Universe is low $l(R)$ should increase as $R$ in line with the postulated $l(r)=l_0 r$ law for a single gravitating body. When the density of the Universe is high and distances between bodies are much smaller than $l_0$ the function of $l(R)$ will be quite different, however. To determine the behavior of $l(R)$ I have conducted a simulation of an expanding Universe containing $10^6$ quantum gravitational sources (1) and sampled the resulting $l(R)$ function. The resulting function $l(R)$ is characterized by the following two limits:

$$l(R) = \begin{cases} l_0 & \text{if } R \to 0; \\ l_0 R & \text{if } R \to \infty. \end{cases} \quad (59)$$

For the magnitude of fluctuations the expected $R^{-1}$ law was confirmed:

$$U_0(R) \propto \frac{1}{R} \quad (60)$$

The equation (59) tells us that the characteristic size of gravitons stays constant until the Universe blows up to a sufficiently large radius (or the mass-density drops below a certain critical value) and then gradually transforms into a linear growth, which becomes an exact linear law when the Universe is sufficiently sparse. The critical radius $R_0$ below which characteristic size of gravitons remains constant regardless of $R$ depends on the choice of the constant $\tilde{l}_0$ in equation (1), which defines the quantum gravitational intrinsic fluctuation frequency.

From the discussion above it follows that $\Lambda \approx$ *const* for small $R$ but the matter density drops as $\rho \propto R^{-3}$ as a result of expansion. Since the matter density drops so rapidly it is reasonable to expect that the early Universe was dominated by cosmological constant, i.e. $\Lambda >> 4G(\rho + p)$ assuming that the initial matter pressure $p$ was small. Based on these arguments the equation (58) for the early Universe can be rewritten as

$$\frac{\ddot{a}}{a} \approx \frac{\Lambda}{3} = \frac{\Lambda_0}{3l_0}(1 - D) \quad (61)$$

Since $1 - D \approx \frac{\gamma^2}{\gamma^2 + v^2} \to 1$ when $\gamma^2 >> v^2$ (i.e. under the assumption of the low initial pressure), the solution to the equation (61) is exponential expansion:

$$a(\tau) = a_0 e^{(\Lambda_0/3l_0)^{1/2}\tau} \quad (62)$$

Thus in the early Universe gravitational energy fluctuations act as an *inflaton* field, which is ultimately responsible for the exponential inflation. According to inflation theories such exponential inflation is necessary to produce flat and homogeneous Universe free of topological defects.

Note that when the Universe becomes sufficiently large $l(R) \propto R$ and the rate of expansion slows down from exponential to a $\tau^2$ power law expansion:

$$\frac{\ddot{a}}{a} = \frac{\Lambda_0}{3l_0 a} \quad (63)$$

$$a(\tau) = \frac{\Lambda_0}{6l_0}\tau^2 + a_0\tau + a_1 \quad (64)$$

Thus the inflationary expansion epoch will come to an end completely when $\Lambda$ drops below $4G(\rho + p)$. This transition will occur in part because of the growth of particle kinetic energy

(i.e. matter pressure) and because of the $R^{-1}$ decrease in the graviton energy $U_0$ – recall the equation (10) stating that $\gamma \propto U_0$, which is equivalent to $\gamma \propto R^{-1}$. Thus as the radius of the Universe $R$ increases with time the assumption that $\gamma \gg v^2$ will be violated and replaced with another limit of $\gamma \ll v^2$. At the transition the tunneling probability will cease to be in the neighborhood of 1 and will start slipping towards zero rapidly. Therefore in the grand scheme of things after a brief period of exponential inflation followed by a long period of power law inflation will come an epoch of steady expansion characterized by the relaxation of $\Lambda$ down to zero:

$$\lim_{R \to \infty} \Lambda = \lim_{R \to \infty} \left( \frac{\Lambda_0}{l_0 R} \frac{\gamma^2}{\gamma^2 + v^2} \right) = 0 \qquad (65)$$

And the rate of expansion is

$$a(\tau) \approx a_0 \tau + a_1 \qquad (66)$$

Note that $\gamma \propto U_0(R) l(R) \propto R^{-1}$ during the period of exponential inflation and $\gamma \approx const$ afterwards. Thus the decrease in the value of $\Lambda$ is caused both by rapid decrease in the energy of the gravitons $U_0$ and by the increase in the kinetic energy of expanding matter.

Observationally non-zero cosmological constant was detected in the recent high-red shift supernovae studies (e.g. Perlmutter et al. 1998; for an overview see Filippenko 2003) and in the WMAP analysis of the CMB anisotropy (for an overview see Lahanas, Mavromatos & Nanopoulos 2003). Theoretically the prediction of quantum gravity that initially significant cosmological constant is currently relaxing to zero at least as fast as fast as $R^{-1}$ is favorable for cosmological inflation theories as it provides a natural mechanism for the exit from the inflationary epoch.

Lastly, in a counterpoint to MOND I would like to mention that MOND fails to make unambiguous and consistent predictions on cosmological scale mostly due to the difficulty of selecting relevant acceleration frames (Scott et. al. 2001), whereas quantum gravity predicts exponential / power law inflation in early epoch and non-zero yet vanishing cosmological constant at present time consistent with the expectations of the cosmological inflation paradigm.

## 13. SUMMARY AND CONCLUSION

A theory of quantum gravity is proposed in which gravitational energy fluctuations – or gravitons – play key role in stellar dynamics by exerting a complimentary graviton scattering force in addition to Newtonian gravity. The additional force is small and is generally significant in the limit of low orbital accelerations. The latter observation results in certain similarity between the proposed the theory of quantum gravity and Milgrom's MOdified Newtonian Dynamics (MOND). Unlike MOND, however, the proposed theory quantum gravity does not suffer from the violation of cornerstone Newton Laws and avoids the difficulty in X-ray cluster cores with where MOND still requires unseen matter and fails to account for flat temperature profile of the X-ray emitting gas. Similarly to MOND, the proposed theory of quantum gravity successfully accounts for *M/L* ratios of various stellar systems including dwarf spheroid galaxies, dwarf, ordinary and giant elliptic galaxies, and galaxy clusters. Although a postulates in MOND, flat rotational curves of spiral galaxies, Tully-Fisher and Faber-Jackson laws follow from the base principles of quantum gravity. Also, the proposed theory of quantum gravity successfully explains the fundamental plane of elliptic galaxies, which can not be described by a virial theorem applied with the accepted CDM model without requiring ad hoc systematic effects, as well as scaling relations for X-ray clusters, which are not consistent with the principle of self-similarity arising from the standard Newtonian gravity / CDM scenario. Lastly, primordial inflation and vanishing but non-zero cosmological constant – or dark energy – can be traced as effects of the universal graviton scattering force induced by expanding matter.

It is remarkable that such a wide variety of phenomena can be derived from two very basic principle of matter particles tunneling through / scattering from the gravitational energy fluctuations. With this encouraging results I call for further investigation of properties of the graviton field and for detailed quantitative analysis of the effects of quantum gravity.

ACKNOWLEDGMENTS

I thank Moti Milgrom for kindly maintaining an email correspondence with me, which although as brief as it was enabled me to grasp the breadth of success and nature of difficulties of MOND, understanding of which was essential for shaping the theory of quantum gravity. I also would like to thank Joe Hershberger, George Fomitchev and Veronica Grigorashvily for helpful discussion and support in preparing this paper.